# Double tunable metamaterial perfect absorber

MAHYAR RADAK,[1,*] SAEED MIRZANEJHAD,[2]

[1]*Department of Atomic and Molecular Physics, University of Mazandaran, Babolsar, Iran*
[2]*Department of Atomic and Molecular Physics, University of Mazandaran, Babolsar, Iran*
**m.radak03@umail.umz.ac.ir*

**Abstract:** This paper focuses on the simulation of a tunable metamaterial absorber designed for the infrared region. Adsorbents offer three different mechanisms to adjust their absorption characteristics. The first method involves changes in temperature. When the temperature changes, the electrical conductivity of the active material, vanadium dioxide (VO2), also changes. This transition between insulating and conducting phases at different temperatures provides the possibility to adjust the absorption spectrum of the metamaterial. By exploiting the thermochromic properties of VO2, the conductivity of the material can be dynamically adjusted over a wide range. Through temperature control, the conductivity of VO2 changes from 200 S/m to $2\times10^5$ S/m, resulting in a continuous adjustment of the absorption peak intensity. The second method relies on applying an electric field. By applying an electric field in the transverse direction of the lead zirconate titanate (PZT) material, the size of the piezoelectric material undergoes changes and bending. Consequently, the entire structure bends, leading to a change in the shape of the absorber. This mechanism allows for precise control of the absorber's shape by applying different voltages, which in turn enables the modification of the absorption peak. This change in distance leads to tunability in the absorption spectrum. The third method combines the effects of temperature and applied electric field. By simultaneously varying the temperature and applying an electric field, the absorption spectrum can be tuned using both factors simultaneously. Through simulation, the tunable metamaterial absorber demonstrates its ability to adjust the absorption properties in different regions of interest, particularly in the infrared region. The proposed scheme offers flexibility and controllability, making it suitable for applications that require tunable absorption in this spectral range.

## 1. Introduction

Metamaterials are a class of artificially engineered materials that have unusual properties that cannot be found in natural materials. They are designed to have properties that can be manipulated at the micrometer and nanometer scale, enabling them to achieve unique physical properties. These materials have attracted attention in recent years due to their potential applications in various fields including optics, acoustics, electromagnetics and mechanics [1-3]. Metamaterials are among the left-handed materials (LH) for which $\varepsilon < 0$ and $\mu < 0$ and with the equation $n = \sqrt{(\varepsilon\mu)}$ the refractive index of negative value $n < 0$ is obtained [4-5]. Also Metamaterials are also used in making materials with near-zero refractive index (NZI) different forms with $\varepsilon \to 0$ constant near-zero transmittance (ENZ) [6-7] and $\mu \to 0$ permeability close to zero (MNZ) [8] or $\mu \to 0$ and $\varepsilon \to 0$ (EMNZ) [9]. Metamaterials are used in the invisibility and coating [10-11]. The perfect metamaterial absorber was first proposed by Landy and his colleagues in 2008 [12] and in the following years, various structures for the perfect absorber such as the L-shaped dual-band [13], T-shaped double-band symmetric and full multi-absorbers with different structures [14] were proposed. Also, recently, nanoscale metamaterials have been designed and simulated, which are complete absorbers in optical regions [15]. Also, Hu Tao and colleagues designed a metamaterial absorber that achieves a high absorption of 0.97 at 1.6 THz [16]. Metamaterial absorbers at microwave and terahertz frequencies have been experimentally demonstrated [17-19]. Thermal tunability relies on temperature-responsive materials [20-21], such as VO2 [22-23] and Germanium-Antimony-Telluride (GST) [24-25].

These materials undergo sudden changes in chemical structure (crystalline to amorphous) or material state (dielectric to metal) in response to temperature. They can be used as part of a layer or as an entire layer in metamaterial with additional processing. Special polymers [26], distilled water [27] and liquid crystals [28] have also been used in thermally tunable metamaterials. Another way to tune absorption is mechanical tunability, which increases the dependence of metamaterial properties on their shape or arrangement. Metamaterials can be fabricated on elastic substrates that can be stretched to produce changes in their properties. Microelectromechanical systems (MEMS) can also be combined to directly change the shape of the metamaterial when a voltage is applied, thereby tuning its properties [29]. Physical deformation using piezoelectric is another method to achieve mechanical adjustability in metamaterial [30]. Specifically, piezoelectricity can be employed to alter the size of the optical cavity in a tunable MM [31].

## 2. Theoretical analysis and simulation

Authors The Drude model is employed to describe the optical characteristics of VO2 in the THz range which can be expressed as [13]

$$\varepsilon(\omega) = \varepsilon_\infty - \frac{\omega_p^2}{\omega(\omega + i\omega_c)} \quad (1)$$

where $\varepsilon_\infty = 12$ is dielectric permittivity at high frequency and $\gamma = 5.75 \times 10^{13}$ rad/s is collision frequency, respectively. The relationship between the plasma frequency $\omega_p$ and conductivity $\sigma$ can be described as

$$\omega_p^2(\sigma) = \frac{\sigma}{\sigma_0} \omega_p^2(\sigma_0) \quad (2)$$

with $\sigma_0 = 3 \times 10^5$ S/m and $\omega_p(\sigma_0) = 1.4 \times 10^{15}$ rad/s. It is assumed that the conductivity of VO2 changes from 200 S/m to $2 \times 10^5$ S/m when it turns the insulator phase into the metal phase [29-30].

The absorption and reflection coefficient $A(\omega)$ is also calculated from the following equation

$$A(\omega) = 1 - |S_{11}(\omega)|^2 - |S_{21}(\omega)|^2 \quad (3)$$

where $S_{11}$ and $S_{21}$ are reflection parameter and transmission parameter, respectively [35].

As an effective medium, metamaterials can be characterized by complex electrical permittivity $\varepsilon(\omega) = \varepsilon_1(\omega) + i\varepsilon_2(\omega)$ and magnetic permeability $\mu(\omega) = \mu_1(\omega) + i\mu_2(\omega)$. Impedance matching theory can also be used to explain the phenomenon of complete absorption of the adsorbent. Here, the effective impedance of the perfect metamaterial absorber (MPA) can be recovered from $\mu$ and $\varepsilon$ can be described as

$$z = \sqrt{\mu/\varepsilon} = \sqrt{\frac{(1+S_{11}^2)-S_{21}^2}{(1-S_{11}^2)-S_{21}^2}} \quad (4)$$

where $\mu$ and $\varepsilon$ represent effective permeability and effective permeability, respectively [29].

Polyamide is a type of polymer material that is highly flexible and possesses excellent mechanical strength [36]. This flexibility is particularly advantageous for applications where the absorber needs to be wrapped around or integrated into curved or non-planar objects. Moreover, the flexibility of polyamide allows for safe integration with piezoelectric materials, such as adjusting the size of the PZT material, without causing damage to the absorber [37]. Additionally, polyamide exhibits resistance to high temperatures, enabling it to withstand elevated temperatures without significant degradation. This thermal stability is crucial for infrared absorbers, as they may be subjected to intense radiation or environmental conditions that generate heat. The ability of polyamide to maintain its integrity even during temperature changes can be utilized to tune the absorption properties through a VO2 phase changing. Furthermore, polyamide is compatible with standard manufacturing techniques, including spin coating, photolithography, and etching [38-39]. This ease of processing and fabrication facilitates efficient production and allows for precise control over the design and dimensions of the absorber. In conclusion, polyamide is a highly desirable material for metamaterial absorbers due to its flexibility, mechanical strength, thermal stability, and compatibility with common manufacturing processes.

The metamaterial consists of a rectangular unit with a length of 1 µm and a width of 0.3 µm. The structure begins with a thin gold layer of 0.03 µm thickness, followed by a layer of vanadium dioxide with a thickness of 0.05 µm, and the next layer is the polyamide layer and PZT, so that the PZT layer with a size of 0.35 micrometer is placed between the two polyamide layers on the left with a length of 0.55 micrometer and the right side with a length of 0.1 micrometer. The subsequent layer consists of vanadium dioxide with a thickness of 0.1 µm. On top of this vanadium dioxide layer, three gold double ring structures are placed. The larger ring has an outer radius of 0.1 µm and an inner radius of 0.07 µm. The smaller ring has an outer radius of 0.05 µm and an inner radius of 0.03 µm. These rings are positioned at a distance of 0.1 µm from each other. The detailed design and arrangement of the layers and structures offer opportunities for further exploration and investigation of their electromagnetic properties and performance characteristics.

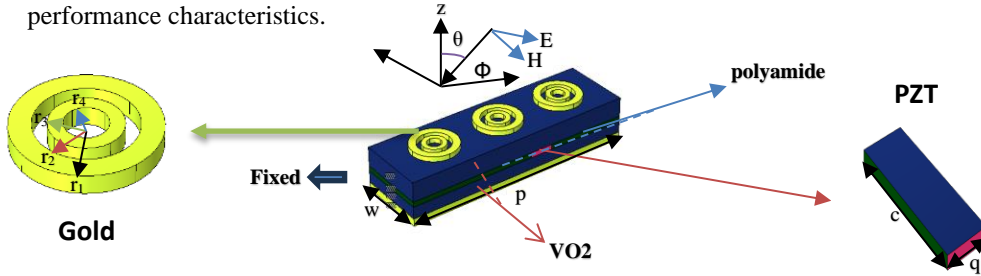

Fig. 1. The proposed metamaterial absorber structure features a rectangular unit with a length (p) of 1 µm and a width (w) of 0.3 µm. The design incorporates three double gold rings with radii: $r_1 = 0.1$ µm, $r_2 = 0.07$ µm, $r_3 = 0.05$ µm, and $r_4 = 0.03$ µm. Additionally, the structure includes a piezoelectric material (PZT), which has a length (c) of 0.35 µm and a width (q) of 0.3 µm.

## 3. Results

Figure (1) shows the designed ultra-wide band infrared (IR) absorber unit cell. This unit cell design serves as the building block for the complete absorber structure, which can be replicated to form an array to improve performance.

### 3.1. Tunability based on VO2 conductivity change

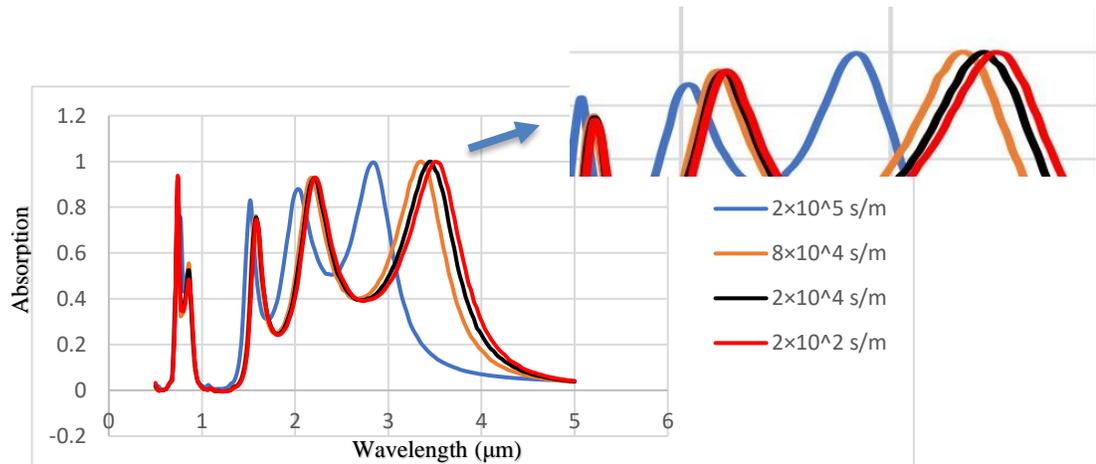

Fig. 2. Absorption spectrum of metamaterial adsorbent, effect of changes in VO2 conductivity at $\varphi = 0°$ and $\theta = 45°$.

As shown in Figure 2, we observed notable variations in the resonance frequencies of the absorber corresponding to different conductivity values. Specifically, we found that at a conductivity of $2\times10^5$ s/m, the resonance frequency of the metamaterial absorber was measured at 106.76 THz. As the conductivity decreased to $8\times10^4$ s/m, the resonance frequency shifted to 89.49 THz. Similarly, for conductivities of $2\times10^4$ s/m and $2\times10^2$ s/m, the resonance frequencies were determined to be 87.21 THz and 85.65 THz, respectively. Furthermore, our analysis revealed that regardless of the conductivity value, the metamaterial absorber exhibited peak absorption values exceeding 99.99%. This remarkable absorption performance across the different conductivity ranges underscores the effectiveness and robustness of the proposed design. These findings highlight the potential of utilizing temperature-induced conductivity changes in vanadium oxide to achieve tunable absorption characteristics in metamaterial structures. In addition to the absorption peak at the corresponding resonance frequency, our absorber also shows three other absorption peaks, which for a conductivity of $2\times10^5$ S/m, the absorber exhibited absorption percentages of 88% at 146 THz, 82% at 197.28 THz, and 76% at 389.11 THz. Similarly, at a conductivity of $8\times10^4$ S/m, absorption percentages of 92% at 137.46 THz, 76% at 189.83 THz, and 83% at 405.13 THz were observed. At a conductivity of $2\times10^4$ S/m, the absorber demonstrated absorption percentages of 92% at 135.51 THz, 75% at 189.24 THz, and 93% at 405.13 THz. Lastly, at a conductivity of $2\times10^2$ S/m, absorption percentages of 93% at 135 THz, 74% at 189.24 THz, and 93% at 405 THz were measured.

### *3.2. Tunability based on piezoelectric voltage change*

By applying voltage to the PZT element, strategically positioned within the metamaterial structure, we induce controlled deformation, leading to remarkable changes in the overall absorber shape. As the voltage is applied to the PZT element, the controlled bending deformation occurs, causing a notable change in the absorber's shape. Consequently, this voltage-induced deformation enables precise modulation of the resonance absorption peaks of our metamaterial absorber.

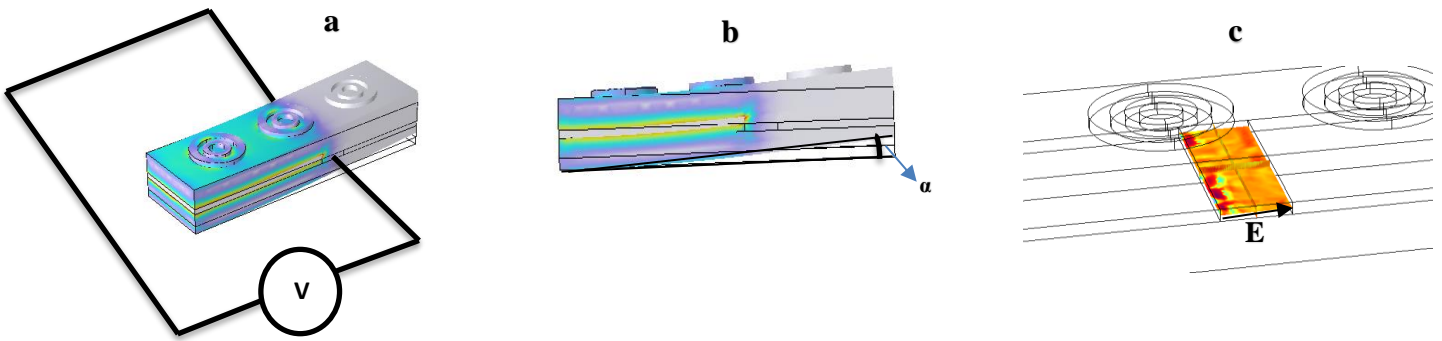

Fig. 3. The Effects of Applied Voltage on Absorber Bending: (a) Side View after Applying Voltage: This view showcases the absorber's response to applied voltage, depicting the bending that occurs as a result. (b) Horizontal View and Bending of the Absorber: This perspective provides a horizontal view of the absorber, highlighting the degree of bending that occurs. The bending is quantified using the alpha angle, which indicates the extent of the absorber's deformation. (c) Applying Voltage to Create a Field within the Piezoelectric Material: This section demonstrates the application of voltage to two points, generating an electric field within the piezoelectric material. The electric field induces changes in the material's shape, leading to the alteration of the absorber's overall structure.

According to Figure 3: after applying voltage to the piezoelectric material (PZT) in the metamaterial, the bending phenomenon occurs due to the piezoelectric effect. The piezoelectric effect is a property exhibited by certain materials, including PZT, wherein they generate electric charges or a voltage when subjected to mechanical stress or strain. In the context of the metamaterial absorber, when voltage is applied to the PZT layer, an electric field is created within the material. This electric field induces strain or deformation in the PZT layer, causing it to expand or contract. Since the PZT layer is integrated into the absorber structure, this expansion or contraction of the PZT layer leads to a corresponding change in the overall shape of the absorber. The bending of the absorber is a result of the mechanical coupling between the PZT layer and the surrounding materials. As the PZT layer deforms, it exerts mechanical forces on the adjacent layers of the absorber, causing them to experience stress and resulting in a bending or curving motion. Additionally, this work presents a significant advantage as it allows for precise control of the amount of absorption by regulating the application of voltage to the PZT layer. By adjusting the voltage level, the degree of bending and deformation in the metamaterial absorber can be finely tuned. This level of control directly influences the absorption characteristics of the material, enabling the manipulation of its absorption properties in real-time.

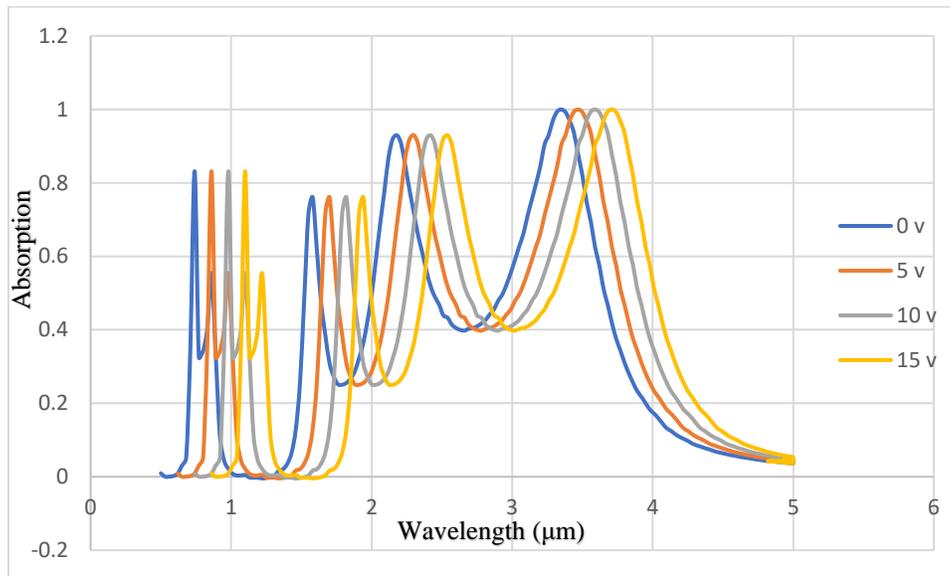

Fig. 4. Controlling the absorption spectrum of the metamaterial by applying different voltages to PZT in a state where the conductivity of vanadium oxide in s/m is $2\times10^5$ s/m.

*3.3. Tunability based on the combination of two variables, voltage change and conductivity changes*

In this part, we will show that if we control the temperature at the same time and at the same time if we apply the voltage to the absorber, we can control the absorption spectrum. This synergistic approach capitalizes on the interplay between thermal effects and electrical stimuli, creating an avenue for fine-tuned spectral manipulation. The integration of temperature control enhances the versatility and precision of the voltage-induced modulation, expanding our ability to tailor the absorption characteristics of metamaterials.

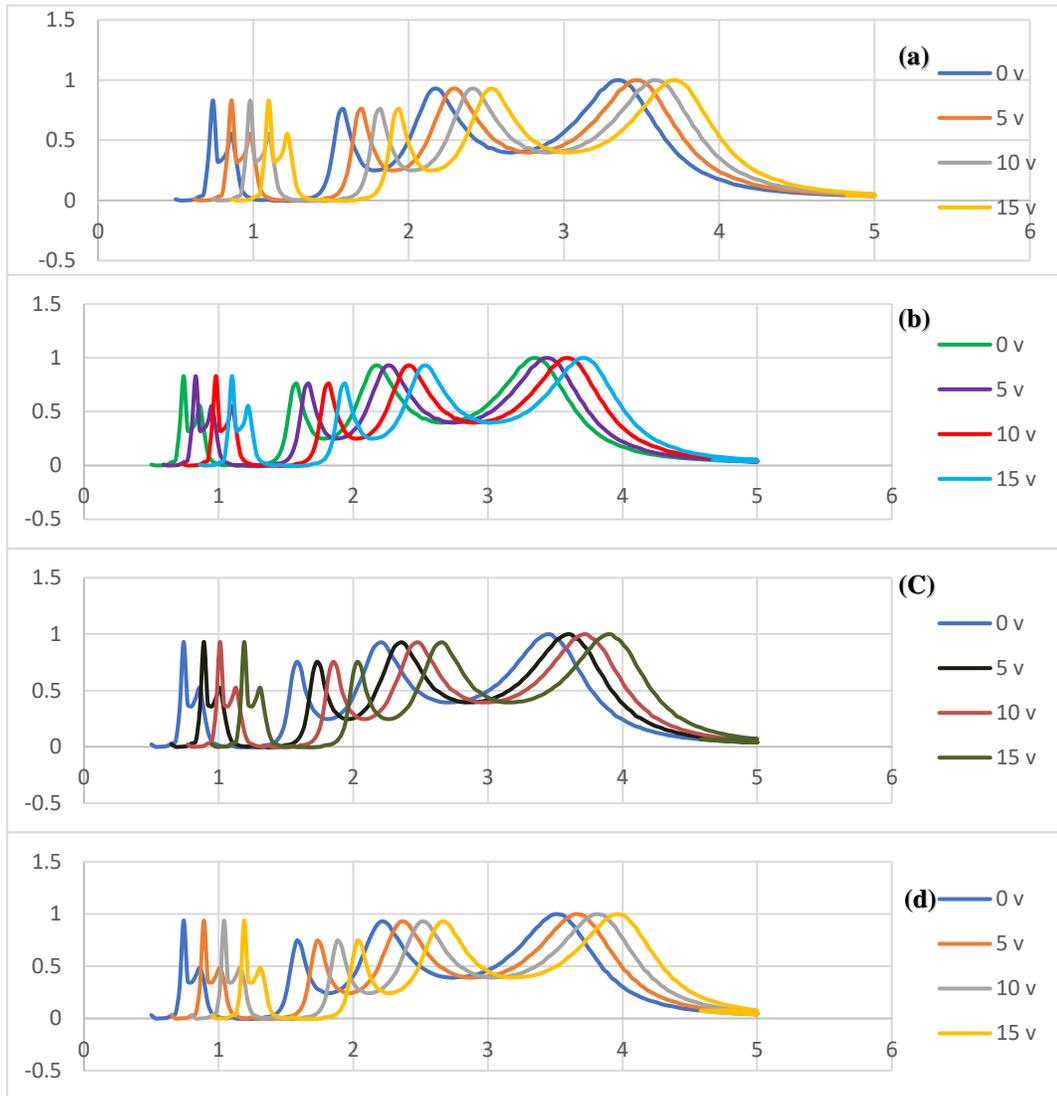

Fig. 5. Controlling the absorption spectrum of the metamaterial both by applying different voltages to PZT and applying conductivity changes due to temperature changes in VO2. Conductivity of vanadium oxide (a) $2\times10^5$ s/m, (b) $8\times10^4$ s/m, (c) $2\times10^4$ s/m and (d) $2\times10^2$ s/m.

## 4. Discussions

In this section, we will explore the impact of absorption on the angle of incidence and azimuthal angle, ranging from 0 to 45 degrees. Additionally, we will investigate how the radius of the outer ring (r1), the radius of the inner ring (r3), and the thickness of the double rings affect the absorption characteristics of the metamaterial. By examining absorption at different angles of incidence and azimuthal angles, we can gain insights into how the metamaterial interacts with incident light from various directions. This analysis will provide valuable information on the suitability of the metamaterial for applications where incident light arrives at different angles.

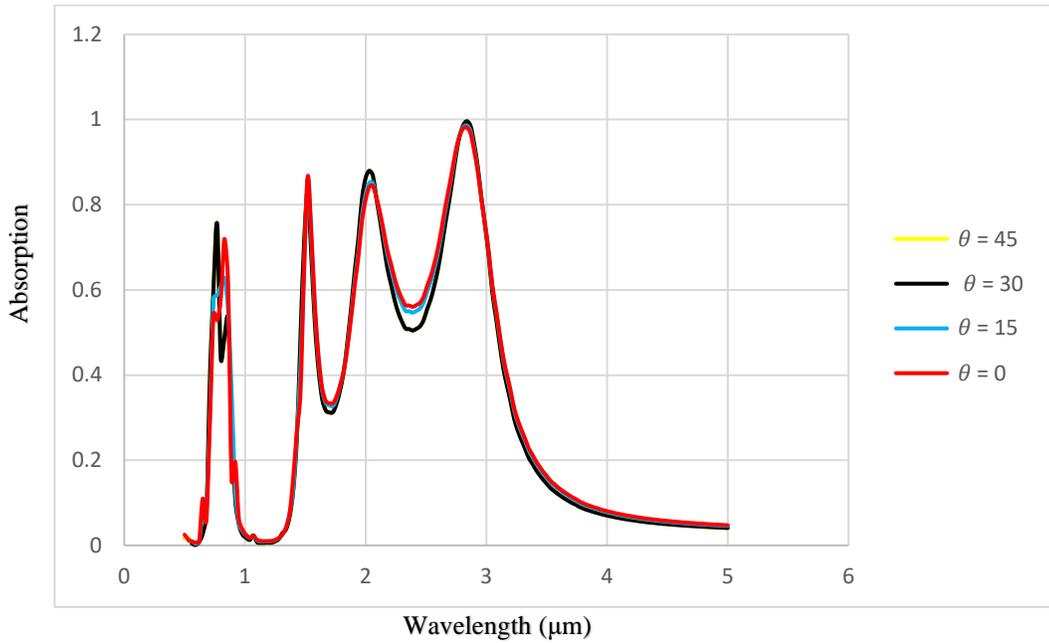

Fig. 6. Absorption spectra of metamaterials at different incident angles θ for TE configuration (φ = 0°, conductivity of vanadium oxide = 2×10$^5$ s/m) and without applying voltage.

In Figure 6, The range of variation of θ is from 0 to 45 degrees, while the angle φ = 0º is fixed.

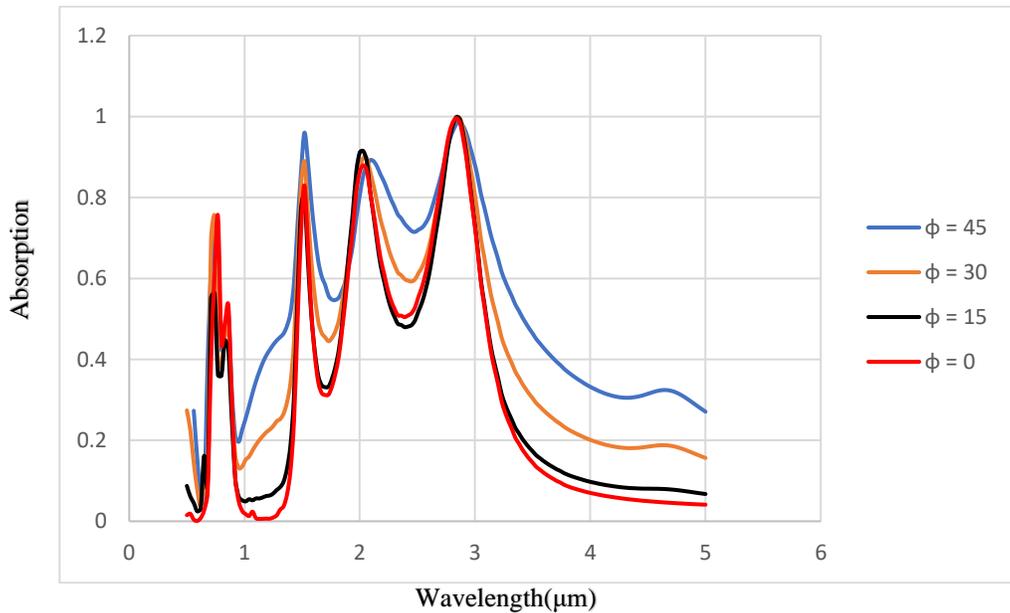

Fig. 7. Absorption spectra of metamaterials at different azimuthal angle φ for TE configuration (θ = 45º, conductivity of vanadium oxide = 2×10$^5$ s/m) and without applying voltage.

In Figure 7, The range of variation of φ is from 0º to 45º, while the angle θ = 45º is fixed.

One of the remarkable features of the proposed metamaterial absorber is its independence from both the incident angle, denoted as $\theta$, and the azimuthal angle, denoted as $\varphi$, across a wide range of 0º to 45º. This means that the absorption peaks of the absorber exhibit minimal changes even when the incident angle $\theta$ and azimuthal angle $\varphi$ are varied within this range. In other words, the absorber maintains its high absorption efficiency consistently across this angular span. This characteristic is particularly advantageous as it ensures consistent performance and effectiveness regardless of the orientation or direction from which the electromagnetic waves impinge on the absorber within the 0º to 45º range. The absorber's absorption peaks exhibit minimal variation with respect to changes in both the incident angle $\theta$ and azimuthal angle $\varphi$, enhancing its versatility and stability. The absorber's ability to maintain its absorption efficiency for incident angles and azimuthal angles ranging from 0 to 45 degrees makes it well-suited for various applications that require reliable and angle-independent absorption in the infrared region within this specific angular range. Whether the waves approach the absorber from different angles or orientations within this range, the absorber consistently delivers optimal performance, providing reliable absorption efficiency. By offering angle independence and stability within the 0º to 45º range, the proposed metamaterial absorber showcases its potential for practical use in a wide range of engineering and scientific applications. Its ability to maintain high absorption efficiency across different incident angles and azimuthal angles within this specific angular span ensures reliable performance in real-world scenarios.

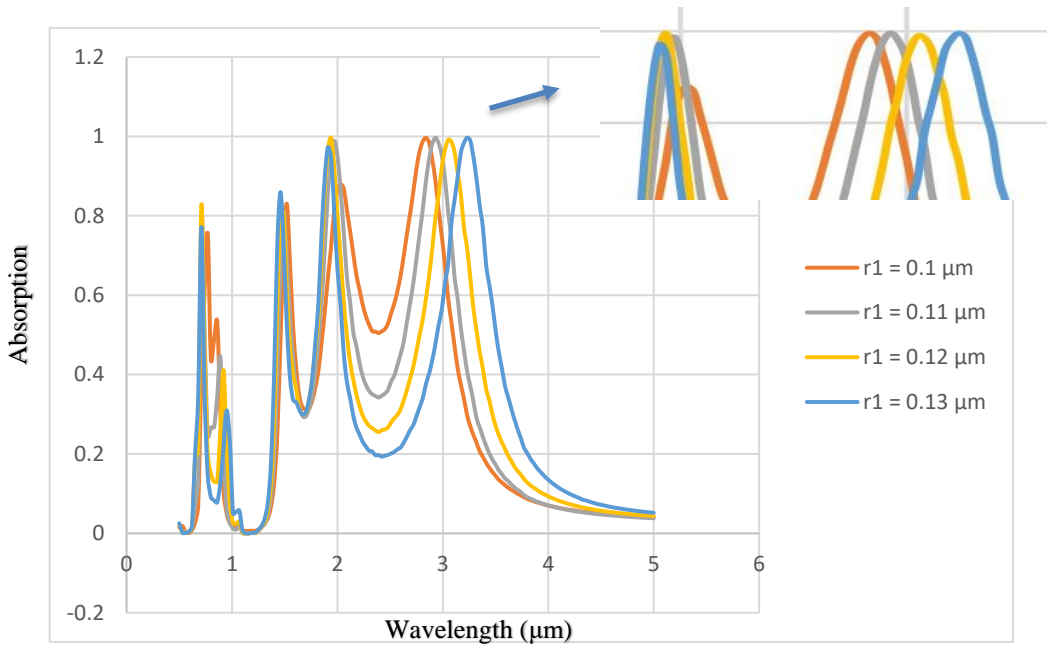

Fig. 8. Exploring the effect of variable outer ring radius (r1) on the absorption spectrum with fixed radius (r2, r3, and r4) in metamaterial structures.

In Figure 8, by altering the external radius of the larger ring (r1), it has been observed that the absorption peaks undergo slight changes. Among these changes, the most favorable outcome occurs when the radius of $r_1$ is set at 0.12 μm. This particular configuration leads to highly efficient absorption at two specific wavelengths, 3.05 μm and 1.93 μm. At a wavelength of 3.05 μm, the absorption rate reaches an impressive 99.1%. This means that nearly all incident IR at this wavelength is absorbed by the double-ring metamaterial structure, resulting in minimal reflection or transmission. This level of absorption suggests a strong resonance between the incident light and the metamaterial's internal properties. Similarly, at a wavelength of 1.93 μm, the absorption rate further increases to 99.6%. This indicates that an even greater proportion of the incident IR at this specific wavelength is absorbed, making it an optimal configuration for applications requiring high absorption efficiency. The observed changes in absorption peaks highlight the sensitivity of the double-ring metamaterial structure to variations in the external radius of the larger ring. By precisely tuning this parameter to 0.12 μm, the metamaterial exhibits remarkable absorption characteristics at the mentioned wavelengths. Specifically, when the size $r_1$ is set at 0.11μm and 0.12 μm, we have observed perfect double-band absorption.

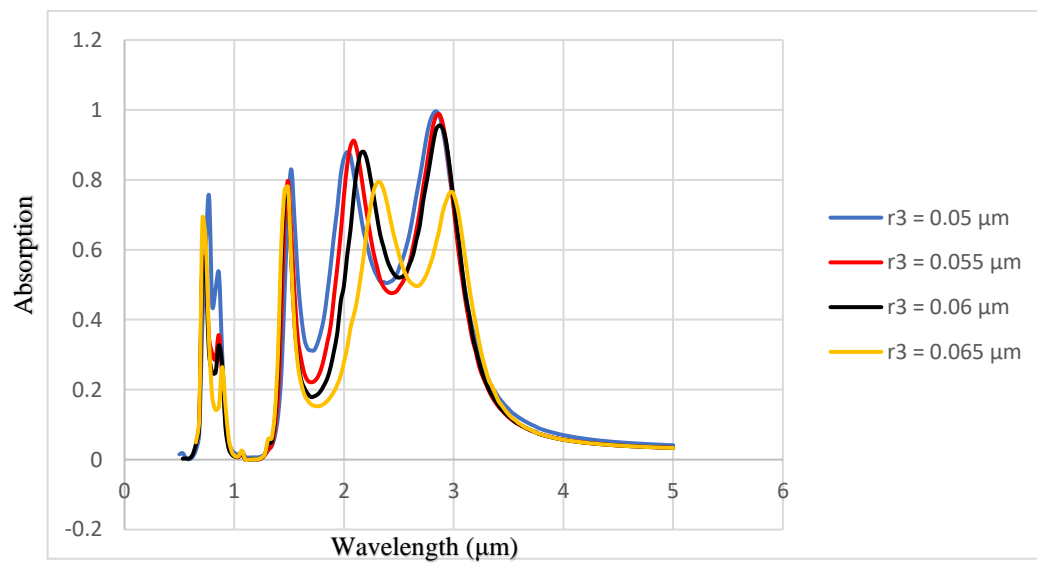

Fig. 9. Exploring the effect of variable outer ring radius (r3) on the absorption spectrum with fixed radius ($r_1$, $r_2$, and $r_4$) in metamaterial structures.

By changing the outer radius of the small ring (r3), we have observed slight changes in the absorption peaks. Among these changes, the most optimal configuration is achieved at an absorption peak of 2.09 μm when the radius r3 is set to 0.055 μm. In this configuration, the absorption reaches 91%. Similarly, for the absorption peak at 2.84 μm, the most optimal radius is 0.05 μm, resulting in a remarkable absorption rate of 99.6%. These observations demonstrate the sensitivity of the absorption peaks to variations in the outer radius of the small ring. Figure 8 and 9, illustrates the alterations in the absorption diagram as the sizes of the inner and outer radii ($r_1$, $r_3$) change.

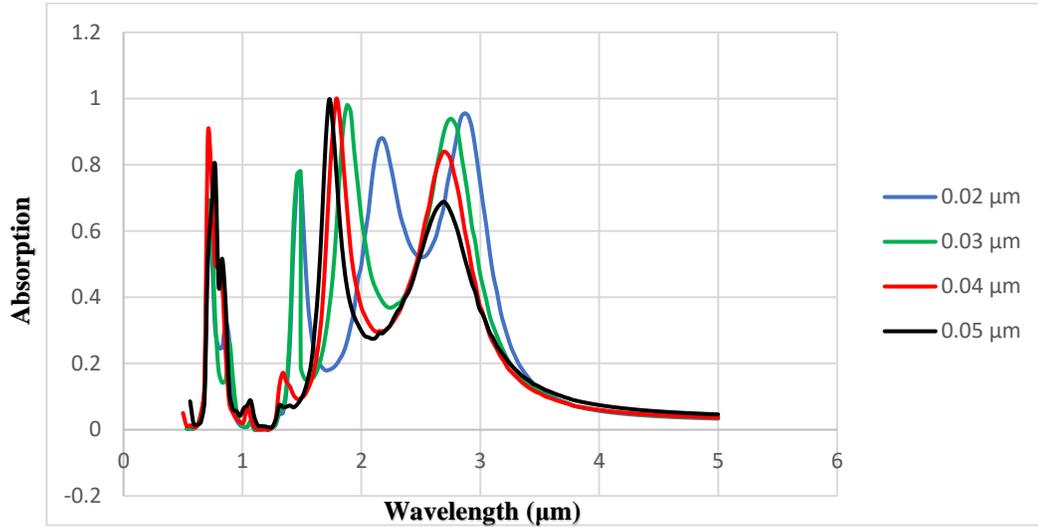

Fig. 10. Exploring the impact of double rings thickness on the absorption spectrum

In Figure 10, we have observed a dependence of the absorption peak on the increasing thickness of the gold double ring. At a thickness of 0.03 μm, two absorption peaks are observed: one at a wavelength of 2.75 μm with 94% absorption, and the other at a wavelength of 1.87 μm with absorption above 98%. Furthermore, for a thickness of 0.04 μm, we have achieved 99.99% absorption at a wavelength of 1.79 μm and 91% absorption for a wavelength of 3.62 μm and for a thickness of 0.05 μm, we have achieved 99.99% absorption at a wavelength of 1.72 μm and 68% absorption for a wavelength of 2.68 μm These observations highlight the impact of thickness variations on the absorption characteristics of the gold double ring structure. Increasing the thickness leads to changes in the absorption peaks, resulting in different absorption strengths at different wavelengths. This dependence on thickness provides an avenue for tailoring and optimizing the absorption properties of the structure for specific applications.

## 5. Conclusions

In this study, we investigated a metamaterial absorber with dual tunability. The absorber was designed with three gold double rings integrated into the active material layer of vanadium dioxide. Our findings demonstrated that by leveraging the conductivity variations of vanadium dioxide induced by temperature changes and applying voltage to the PZT, we could effectively control the absorption area. Remarkably, we showcased the simultaneous utilization of both approaches, combining the conductivity control and piezoelectric effects to achieve enhanced control over the absorption area. In addition, we propose an innovative approach that we are currently studying in this subject. In the future, we aim to report results that include an engineered combination of three active ingredients. This combination would provide a means to simultaneously control the absorption area in eight different ways. This method offers a remarkable level of versatility and opens up exciting possibilities for tailoring and manipulating the absorption properties of metamaterials. Moreover, they propose the possibility of further enhancing absorption capabilities through strategic combinations of the metamaterial structure. This research opens up new avenues for engineered control and manipulation of absorption properties in metamaterials.

*Acknowledgments*

The authors would like to express their gratitude to Mazandaran University for providing the computational resources required for the simulations conducted in this article.

*Disclosures*

The authors declare no conflicts of interest.

*Data availability statement*

Data underlying the results presented in this paper are not publicly available at this time but may be obtained from the authors upon reasonable request.